\documentclass[pra,unsortedaddress,superscriptaddress,a4paper,nofootinbib,%
nobalancelastpage,preprintnumbers]{revtex4}%
\usepackage{epsfig}
\usepackage{graphicx}
\usepackage{graphics}
\usepackage{xspace}
\usepackage{amssymb}
\usepackage{amsmath}
\usepackage{latexsym}
\usepackage{natbib}
\usepackage{mathrsfs}
\usepackage{url}

\def\q{\mbox{\boldmath $q$}}
\def\p{\mbox{\boldmath $p$}}
\def\r{\mbox{\boldmath $r$}}
\def\J{\mbox{\boldmath $J$}}
\def\ss{\mbox{\boldmath $\sigma$}}
\def\tt{\mbox{\boldmath $\tau$}}

\begin{document}

\title{Center-of-mass effects in electromagnetic  two-proton knockout 
reactions}

\author{ Carlotta Giusti}
\affiliation{Dipartimento di Fisica Nucleare e Teorica,
Universit\`a degli Studi di Pavia\\ 
Istituto Nazionale di Fisica Nucleare,
Sezione di Pavia, I-27100 Pavia, Italy}
\author{ Franco Davide Pacati}
\affiliation{Dipartimento di Fisica Nucleare e Teorica,
Universit\`a degli Studi di Pavia\\ 
Istituto Nazionale di Fisica Nucleare,
Sezione di Pavia, I-27100 Pavia, Italy}
\author{Michael Schwamb}
\affiliation{ Dipartimento di Fisica, Universit\`{a} degli Studi di Trento \\
and Istituto Nazionale di Fisica Nucleare,
Gruppo Collegato di Trento,  I-38100 Povo (Trento), Italy }
\affiliation{ European Center for Theoretical Studies in Nuclear
Physics and Related Areas (ECT$^{\ast}$),
I-38100 Villazzano (Trento), Italy}
\author{Sigfrido Boffi}
\affiliation{Dipartimento di Fisica Nucleare e Teorica,
Universit\`a degli Studi di Pavia\\ 
Istituto Nazionale di Fisica Nucleare,
Sezione di Pavia, I-27100 Pavia, Italy}


\begin{abstract}
The role of center-of mass (CM) effects in the one-body nuclear 
current in the description of electromagnetically induced two-nucleon
knockout reactions is discussed in connection with the problem of the lack of 
orthogonality between initial bound states
and final scattering states obtained by the use of an energy-dependent optical 
model potential. Results for the cross sections of 
the exclusive $^{16}$O(e,e$'$pp)$^{14}$C and 
$^{16}$O($\gamma$,pp)$^{14}$C knockout reactions in different kinematics are presented 
and discussed. In super-parallel kinematics CM effects produce a strong 
enhancement of the $^{16}$O(e,e$'$pp)$^{14}$C$_{\rm g.s.}$ cross section which 
strongly reduces the destructive interference between the one-body and 
$\Delta$-current and the sensitivity to the treatment of the $\Delta$-current 
found in previous work. \\

 continuum

\end{abstract}

\maketitle

\section{Introduction}
\label{sec:intro}
The investigation of nuclear structure is one of the most important
and ambitious aims of hadronic physics. A reasonable starting point is
offered by the independent particle shell model. However, the 
 incorporation of additional short-range correlations (SRC)
  beyond a mean-field description turns out to be inevitably necessary
 for a proper description of nuclear binding. The most direct reaction 
to study SRC is naturally electromagnetically induced two-nucleon
knockout. Intuitively, the probability that a photon is absorbed by a
nucleon pair should be a direct measure for SRC. However, due to
competing two-body effects like meson-exchange currents (MEC) or final
state interactions (FSI), this simple picture needs to be modified
 in order to obtain quantitative predictions. Ideally, the role of 
MEC and FSI should be small or at least under control in order  to
extract information on SRC from experiment. This requires a theoretical 
approach which should be as comprehensive as  possible. 
 An overview over  the available theoretical models till the middle
of the 90s can be found in \cite{Ox}.  Presently, different models are 
available
(see \cite{AnC04,RyN04,sch2} and references therein).
Due to the conceptual complexity of the nuclear many-body problem,
various  approximations and simplifying assumptions are needed for practical 
calculations. Thus, usually
 different treatments of initial bound and final scattering 
states are adopted in the models. 
 
 In the Pavia model \cite{sch2,GP} bound and scattering states are, in 
 principle, consistently derived as eigenfunctions of an energy-dependent 
 non-Hermitian Feshbach-type optical potential. However, in actual 
 calculations the initial hadronic state is obtained from a recent calculation 
 of the two-nucleon spectral function \cite{barb} where different types of 
 correlations are included consistently. For the final hadronic state, a 
 complex phenomenological optical potential,  derived through a fit to 
 nucleon-nucleus scattering data, is used for the description of the FSI 
 between the outgoing nucleons and the residual nucleus.
The mutual nucleon-nucleon  interaction (NN-FSI) in the final  state  can be 
taken into account at least perturbatively \cite{sch1a,sch1}.

Independently of the specific prescriptions adopted in the calculations, a 
conceptual problem arises in the model where the initial and final states, 
which are eigenfunctions of an energy-dependent optical potential at different 
energies, are, as such, not orthogonal. Indeed, the process involves 
transitions between bound and continuum states which must be orthogonal, since 
they are eigenfunctions of the full nuclear many-body Hamiltionian at 
different energies.
Orthogonality is in general lost in a model when the descripion is restricted 
to a subspace where other channels are suppressed. The description of direct
knockout reactions in terms of the eigenfunctions of a complex 
energy-dependent optical potential
considers only partially the contribution of competing
inelastic channels. The remaining effects due to occuring inelasticities  
can, in principle, be taken into account by a
 suitable effective transition operator,
which removes the orthogonality defect of the model wave 
functions \cite{ort}. In practice, however, the usual approach does not make 
use of an effective  operator.

The present paper deals with the proper treatment of all the  
CM effects in the matrix elements of the one-body nuclear current
in connection with the problem of the lack of orthogonality between 
initial and final states in the calculation of the cross section of the
electromagnetic two-nucleon knockout reactions.

The reaction mechanism and CM effects are discussed in sect. 
\ref{sec:reaction}.  Different prescriptions are proposed to cure the 
spuriosity which may result in the numerical calculations as a consequence 
of the orthogonality defect. These prescriptions are discussed and related to a 
proper treatment of all the CM effects in the transition matrix elements. 
In sect. \ref{sec:results} the effects of CM and orthogonality are illustrated, 
with specific numerical examples in selected kinematics, for the exclusive 
$^{16}$O(e,e$'$pp)$^{14}$C and  $^{16}$O($\gamma$,pp)$^{14}$C knockout 
reactions. 
A summary and some conclusions can be found in sect. \ref{sec:conclusions}.

\section{Reaction mechanism and center-of-mass effects}
\label{sec:reaction}

The basic ingredients for the calculation of the cross section of the reaction 
induced by a real or virtual photon, with momentum $\q$, where two nucleons, 
with momenta $\p'_{1}$, and $\p'_{2}$, are ejected from a nucleus, are given  
by the transition matrix elements of the charge-current density 
operator between initial and final nuclear states 
\begin{equation}
J^\mu (\q) = \int  
< \Psi_{\mathrm{f}} | \hat{J}^\mu(\r) |
\Psi_{\mathrm{i}} >
{\mathrm{e}}^{\,{\mathrm{i}}{\footnotesize \q} \cdot
{\footnotesize \r}} {\mathrm d}\r.
\label{eq:jm}
\end{equation} 
Bilinear products of these integrals give the components of the 
hadron tensor, whose suitable combinations allow the calculation of all the 
observables available from the reaction process \cite{Ox,GP}.

If the residual nucleus is left in a discrete eigenstate of its Hamiltonian,
i.e. for an exclusive process, and under the assumption of the direct 
knock-out mechanism, the matrix elements of Eq.~(\ref{eq:jm})
can be written as~\cite{GP,GP97}\footnote{Spin/isospin indices are generally 
suppressed in the formulas of this paper for the sake of simplicity.}
\begin{equation}
J^{\mu}({\mbox{\boldmath $q$}}) = \int
{\widetilde \psi}_{\rm{f}}^{*}
({\mbox{\boldmath $r$}}_{1},{\mbox{\boldmath $r$}}_{2})
J^{\mu}({\mbox{\boldmath $r$}},{\mbox{\boldmath $r$}}_{1},
{\mbox{\boldmath $r$}}_{2})
 {\widetilde \psi}_{\rm{i}}
({\mbox{\boldmath $r$}}_{1},{\mbox{\boldmath $r$}}_{2})
{\rm{e}}^{{\rm{i}}
\hbox{\footnotesize {\mbox{\boldmath $q$}}}
\cdot
\hbox{\footnotesize {\mbox{\boldmath $r$}}}
} {\rm d}{\mbox{\boldmath $r$}} \,
{\rm d}{\mbox{\boldmath $r$}}_{1} {\rm d}{\mbox{\boldmath $r$}}_{2}. 
 \label{eq:jq}
\end{equation}

Eq.~(\ref{eq:jq}) contains three main ingredients: the two-nucleon scattering 
wave function ${\widetilde \psi}_{\rm{f}}$, 
the nuclear current $J^{\mu}$ and the 
two-nucleon overlap integral (TOF)  ${\widetilde \psi}_{\rm{i}}$ 
between the ground state of the
target and the final state of the residual nucleus.

The nuclear current $J^{\mu}$ is the sum of a one-body  and a two-body 
contribution, i.e.
\begin{equation}
J^{\mu}({\mbox{\boldmath $r$}},{\mbox{\boldmath $r$}}_{1},
{\mbox{\boldmath $r$}}_{2}) = J^{(1)\mu}({\mbox{\boldmath $r$}},
{\mbox{\boldmath $r$}}_{1}) + J^{(1)\mu}({\mbox{\boldmath $r$}},
{\mbox{\boldmath $r$}}_{2}) +
 J^{(2)\mu}({\mbox{\boldmath $r$}},
{\mbox{\boldmath $r$}}_{1},{\mbox{\boldmath $r$}}_{2}).
\label{eq:jq12}
\end{equation}
The one-body (OB) part includes the longitudinal charge term and the transverse 
convective and spin currents, and can be written as
\begin{equation}
J^{(1)\mu}(\r,\r_k) = j^{(1)\mu}(\r,\ss_k) \, \delta(\r - \r_k)
\label{eq:j1}
\end{equation}
with $k=1,2$.
The two-body current is derived from the effective Lagrangian of \cite{Peccei}, 
performing a non relativistic reduction of the lowest-order Feynman diagrams 
with one-pion exchange. We have thus currents corresponding to the seagull and 
pion-in-flight diagrams, and to the diagrams with intermediate $\Delta$-isobar 
configurations~\cite{gnn}, i.e.
\begin{equation}
\J^{(2)}(\r,\r_{1},\r_{2})  = 
\J^{\mathrm{sea}}(\r,\r_{1},\r_{2}) 
 + \ \J^{\pi}(\r,\r_{1},\r_{2}) 
 +  \J^{\Delta}(\r,\r_{1},\r_{2}) . \label{eq:nc}
\end{equation}
For two-proton emission the seagull and pion-in-flight me\-son-ex\-cha\-nge 
currents and the charge-exchange contribution of the $\Delta$-current are 
vanishing in the nonrelativistic limit. 
The surviving components of the $\Delta$-current can be written as 
\begin{equation}
J^{(2)\mu}(\r,\r_1,\r_2) = j^{(2)\mu}(\r_{12},\ss_2,\tt_2) \, 
\delta(\r - \r_1)  +
j^{(2)\mu}(\r_{12},\ss_1,\tt_1) \, \delta(\r - \r_2).
\label{eq:j2}
\end{equation}
with $\r_{12}=\r_1-\r_2$.
Details of the nuclear current components can be found in 
\cite{sch2,gnn,WiA97,mec}. More specifically, the various treatments and 
parametrizations of the $\Delta$-current used in the calculations are given 
in \cite{sch2}.

In order to evaluate the transition amplitude of Eq.~(\ref{eq:jq}), for the 
three-body system consisting of the two protons, 1 and 2, and of the residual 
nucleus $B$, it appears to be natural to work with  CM coordinates \cite{GP,jack} 
\begin{eqnarray}
&{\mbox{\boldmath $r$}}_{1B}& = {\mbox{\boldmath $r$}}_{1} - 
{\mbox{\boldmath $r$}}_{B}, \,\,{\mbox{\boldmath $r$}}_{2B} = 
{\mbox{\boldmath $r$}}_{2} - {\mbox{\boldmath $r$}}_{B}, \nonumber \\
&{\mbox{\boldmath $r$}}_{B}& = \sum _{i=3} ^{A} {\mbox{\boldmath $r$}}_{i}
/ (A-2).
 \label{eq:cm}
\end{eqnarray}
The conjugated momenta are given by
\begin{eqnarray}
{\mbox{\boldmath $p$}}_{1B} &=& \frac{A-1}{A}
{\mbox{\boldmath $p'$}}_{1} - \frac{1}{A} {\mbox{\boldmath $p'$}}_{2}
- \frac{1}{A} {\mbox{\boldmath $p$}}_{B} \,\, , \\
{\mbox{\boldmath $p$}}_{2B} &=& -\frac{1}{A}
{\mbox{\boldmath $p'$}}_{1} + \frac{A-1}{A} {\mbox{\boldmath $p'$}}_{2}
- \frac{1}{A} {\mbox{\boldmath $p$}}_{B} \,\, , \\
{\mbox{\boldmath $P$}} &=& 
{\mbox{\boldmath $p'$}}_{1} + {\mbox{\boldmath $p'$}}_{2}
+  {\mbox{\boldmath $p$}}_{B} \,\, , \label{eq:mom}
\end{eqnarray}
where  $\p_{B}=\q-\p'_1-\p'_2$ is the momentum of the residual nucleus in the
laboratory frame.

With the help of these relations,  one can cast the transition amplitude
(\ref{eq:jq}) into the following form 
\begin{equation}
 J^{\mu}({\mbox{\boldmath $q$}})  = \int
{ \psi}_{\rm{f}}^{*}({\mbox{\boldmath $r$}}_{1B},
{\mbox{\boldmath $r$}}_{2B})
V^{\mu}({\mbox{\boldmath $r$}}_{1B},{\mbox{\boldmath $r$}}_{2B})
  { \psi}_{\rm{i}}
({\mbox{\boldmath $r$}}_{1B},{\mbox{\boldmath $r$}}_{2B})
{\rm d}{\mbox{\boldmath $r$}}_{1B} {\rm d}{\mbox{\boldmath $r$}}_{2B},  
 \label{eq:jqcm}
\end{equation}
with the definition 
\begin{equation}\label{eq:def1}
 {\psi}_{\rm{i/f}}
({\mbox{\boldmath $r$}}_{1B},{\mbox{\boldmath $r$}}_{2B})
 :=
 {\widetilde{\psi}}_{\rm{i/f}}
({\mbox{\boldmath $r$}}_{1},{\mbox{\boldmath $r$}}_{2})
\end{equation}
and the expression
\begin{equation}
V^{\mu}({\mbox{\boldmath $r$}}_{1B},{\mbox{\boldmath $r$}}_{2B}) =
\exp \left( i {\mbox{\boldmath $q$}} \frac {A-1}{A}{\mbox{\boldmath $r$}}_{1B}
\right) \exp \left(-i {\mbox{\boldmath $q$}} \frac {1}{A}
{\mbox{\boldmath $r$}}_{2B}\right) 
(j^{(1)\mu}({\mbox{\boldmath $r$}_{1B}},\ss_1) +  
j^{(2)\mu}({\mbox{\boldmath $r$}_{12}},\ss_2,\tt_2))  \quad 
 + \left( 1 \leftrightarrow 2 \right) \, . \label{eq:opcm}
\end{equation}

It is generally thought that the contribution of the OB current to
two-nucleon knockout is entirely due to the correlations included in the
two-nucleon wave function. 
In fact, a OB operator cannot affect two particles if they are not 
correlated. 
It can be seen, however, from Eq.~(\ref{eq:opcm}) that in the CM
frame the transition operator becomes a two-body operator even in the case of a 
OB nuclear current. 
Only in the limit $A \rightarrow \infty$ CM effects are neglected and the 
expression in Eq.~(\ref{eq:jqcm}) vanishes for a pure OB current in 
Eq.~(\ref{eq:opcm}) sandwiched 
between orthogonalized single particle (s.p.) wave functions.
This means that, due to this CM effect, for finite nuclei the OB current can 
give a contribution to the cross section of two-particle emission independently 
of correlations. 
This  effect is similar to the one of the effective charges in electromagnetic 
reactions~\cite{ec}.

The matrix elements of Eq.~(\ref{eq:jqcm}) involve bound and scattering states, 
$\psi_{\rm{i}}$ and $\psi_{\rm{f}}$, which are consistently derived  
  from an energy-dependent non-Hermitian Feshbach-type Hamiltonian
for the considered final state of the residual nucleus.
They are eigenfunctions of this Hamiltonian at negative and positive energy 
values \cite{Ox,GP}.  
However, in practice, it is not possible to achieve this consistency and the
treatment of initial and final states proceeds separately with different 
approximations.

The two-nucleon overlap function (TOF) $\psi_{\rm{i}}$ contains information on 
nuclear structure and correlations. Different approaches are used in
\cite{barb,GP97,GPA98,Kadrev}. In the present calculations the TOF is 
obtained as in \cite{barb}, from the most recent calculation of the two-proton 
spectral function of $^{16}$O, where both SRC and long-range 
correlations are included consistently with a two-step procedure. 

In the calculation of the final-state wave function $\psi_{\rm{f}}$ only the 
interaction of each one of the two outgoing nucleons with the residual nucleus 
is included. Therefore, the scattering state is written as the product of two 
uncoupled s.p. distorted wave functions, eigenfunctions of a complex phenomenological 
optical potential which contains a central, a Coulomb, and a spin-orbit term 
\cite{Nad81}.
The effect of the mutual interaction between the two outgoing nucleons has been
studied in \cite{sch1a,sch1,KM} and can in principle be included as in 
\cite{sch1a,sch1}. 

The matrix element of Eq.~(\ref{eq:jqcm}) contains a spurious contribution since 
it does not vanish when the transition operator $V$ is set equal to 1.
This is essentially due to the lack of orthogonality between the initial and
the final state wave functions. In the model the use of an effective nuclear
current operator removes the orthogonality defect besides taking into account 
space truncation effects \cite{Ox,ort}. In the usual approach of 
Eq.~(\ref{eq:jqcm}), however, the effective operator is replaced by the bare 
nuclear current operator.  Thus, it is this replacement which may introduce a
spurious contribution which is not specifically due to the different 
prescriptions adopted in practical calculations, but is already present in  
Eq.~(\ref{eq:jqcm}), where $\psi_{\rm{i}}$ and $\psi_{\rm{f}}$ are 
eigenfunctions of an energy-dependent Feshbach-type Hamiltonian at different 
energies. In the past this spuriousity was cured by subtracting from the 
transition amplitude the contribution of the OB current without correlations 
in the nuclear wave functions. In detail, the expression 
\begin{equation}
   \int
\psi_{\rm{f}}^{*}({\mbox{\boldmath $r$}}_{1B},{\mbox{\boldmath $r$}}_{2B})
\left( 
\exp \left( i {\mbox{\boldmath $q$}} \frac {A-1}{A}{\mbox{\boldmath $r$}}_{1B}
\right) \exp \left(-i {\mbox{\boldmath $q$}} \frac {1}{A}
{\mbox{\boldmath $r$}}_{2B}\right)  
j^{(1)\mu}({\mbox{\boldmath $r$}_{1B}},\ss_1) +  
 1 \leftrightarrow 2  \right)
 \psi_{\rm{i},no\, Cor}
({\mbox{\boldmath $r$}}_{1B},{\mbox{\boldmath $r$}}_{2B})
{\rm d}{\mbox{\boldmath $r$}}_{1B} {\rm d}{\mbox{\boldmath $r$}}_{2B}
\label{eq:jqcm2}
\end{equation}
was subtracted from  (\ref{eq:jqcm}), where  in the initial state
 $\psi_{\rm{i},no\, Cor}$ SRC are ignored. This
prescription  is denoted as approach A in the proceeding discussions.
 In this approach, however, we do not subtract only the
spuriosity, but also the CM effect given by the two-body operator 
in Eq.~(\ref{eq:opcm}), which is present in the OB current 
independently of correlations  and which is not spurious. 
The relevance of this effect  can be estimated comparing our previous results
with the results of a different prescription, that is denoted as 
approach B in the proceeding discussions, where we subtract  
from  (\ref{eq:jqcm}) instead of  (\ref{eq:jqcm2}) the spurious contribution 
due to the OB current without correlations and without CM corrections.
 This can be achieved by putting the limit $A \rightarrow \infty$ in
 (\ref{eq:jqcm2}), i.e.\ by the expression
\begin{equation}
   \int
\psi_{\rm{f}}^{*}({\mbox{\boldmath $r$}}_{1B},{\mbox{\boldmath $r$}}_{2B})
\left( 
\exp \left( i {\mbox{\boldmath $q$}}{\mbox{\boldmath $r$}}_{1B}
\right)   
j^{(1)\mu}({\mbox{\boldmath $r$}_{1B}},\ss_1) +  
 1 \leftrightarrow 2  \right) 
 \psi_{\rm{i},no\, Cor}
({\mbox{\boldmath $r$}}_{1B},{\mbox{\boldmath $r$}}_{2B})
{\rm d}{\mbox{\boldmath $r$}}_{1B} {\rm d}{\mbox{\boldmath $r$}}_{2B}\,\, .
 \label{eq:jqcm3}
\end{equation}
This prescription gives an improved, although still rough, evaluation of the 
spurious contribution. 

An alternative and more accurate procedure to get rid of the spuriosity is to 
enforce orthogonality between the initial and final states by means of a 
Gram-Schmidt orthogonalization \cite{ortho}. In this approach each one of 
the two s.p. distorted wave functions is orthogonalized to all the s.p. 
shell-model wave functions that are used to calculate the TOF, i.e., for the 
TOF of \cite{barb}, to the h.o. states of the basis used in the calculation of 
the spectral function, which range from  the $0s$ up to the $1p$-$0f$ shell.
This more accurate procedure, that we denote as approach C in the
proceeding discussions, allows us 
to get rid of the spurious contribution to two-nucleon emission due to a 
OB operator acting on either nucleon of an 
uncorrelated pair, which is due to the lack of orthogonality between the s.p. 
bound and scattering states of the pair. In this approach, in consequence,
 no OB current contribution without correlations like (\ref{eq:jqcm2})
 or (\ref{eq:jqcm3}) needs to be subtracted.
Moreover, it allows us to 
include automatically all  CM effects via (\ref{eq:opcm}).

\section{Results}
\label{sec:results}

The effects of CM and orthogonalization have been investigated for the 
exclusive $^{16}$O(e,e$'$pp)$^{14}$C and $^{16}$O($\gamma$,pp)$^{14}$C 
reactions. 

Calculations performed in different situations indicate that the 
results depend on kinematics and on the prescriptions adopted to treat the
theoretical ingredients of the model. The contribution due to the CM effects 
in the OB current without correlations, that were neglected in our previous 
calculations, are in general non negligible. Although in many situations this 
contribution is small and does not change significantly the results, there are
also situations where it is large and produces important quantitative and
qualitative differences. This is the case of the super-parallel kinematics, 
where these effects are maximized. The super-parallel kinematics is 
therefore of particular interest for our study.

In the so-called super-parallel kinematics the two nucleons are ejected 
parallel and anti-parallel to the momentum transfer and, for a fixed
value of the energy $\omega$ and momentum transfer $q$,  it is possible to 
explore, for different values of the kinetic energies of the outgoing nucleons, 
all possible values of the recoil momentum $p_B$. 
This kinematical setting has been widely investigated in our previous work 
\cite{sch2,GP,barb,sch1a,sch1,GP97,GPA98} and is of particular interest from 
the experimental point of view, since it has been realized in the recent 
$^{16}$O(e,e$'$pp)$^{14}$C  \cite{Rosner} and  $^{16}$O(e,e$'$pn)$^{14}$N 
\cite{middleton} experiments at MAMI.
The super-parallel kinematics chosen for the present calculations of the 
$^{16}$O(e,e$'$pp)$^{14}$C reaction is the same already considered in our 
previous work and realized in the experiment \cite{Rosner}
  at MAMI, i.e. the incident
electron energy is $E_{0}=855$ MeV, $\omega=215$ MeV, and $q=316$ MeV/$c$. 

\begin{figure}
\centerline{
\resizebox{0.55\textwidth}{!}{
  \includegraphics{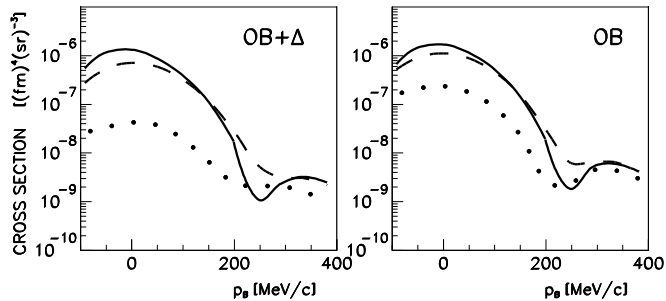}}
}
\caption{
The differential cross section of the $^{16}$O(e,e$'$pp)$^{14}$C reaction to 
the $0^+$ ground state of $^{14}$C as a function of $p_B$ in a 
super-parallel kinematics with $E_0= 855$ MeV, electron scattering angle 
$\theta_e = 18^{\circ}$, $\omega=215$ MeV, and  $q=316$ MeV/$c$.
Different values of $p_B$ are obtained changing the kinetic energies of 
the outgoing nucleons. Positive (negative) values of $p_B$ refer to 
situations where $\p_B$ is parallel (anti-parallel) to $\q$. 
The final results given by sum of the one-body and $\Delta$-currents 
(OB+$\Delta$) are displayed in the left panel, the separate contribution of the 
one-body (OB) current is shown in the right panel. The TOF from the two-proton
spectral function of \cite{barb} and the $\Delta$(NN) parametrization 
\cite{sch2} for the 
$\Delta$-current are used in the calculations. The dotted lines give the 
results of \cite{sch2}, i.e.\ of approach A. The dashed and solid lines
 refer to approach B (improved treatment of the CM contribution 
of the OB current) and approach C (explicit orthogonalization of s.p. bound and
scattering states), respectively.
}
\label{cm1}
\end{figure}
The cross section of the $^{16}$O(e,e$'$pp)$^{14}$C reaction to the $0^+$ 
ground state of $^{14}$C calculated in the super-parallel kinematics is
displayed in Fig.~\ref{cm1} for the three different approaches A (dotted),
B (dashed) and C (solid).
The CM contribution included in approach B produces a large enhancement of 
the cross section calculated with the OB current.  
 The results are shown in the right
panel of the figure, where it can be seen that the enhancement is large for 
recoil momentum values up to about 300 MeV/$c$ and is a factor of about 5 in 
the maximum region. A similar result is obtained with orthogonalized initial 
and final states. In this case the OB cross section is a bit larger at low 
values of the recoil momentum and a bit lower at larger values of 
$p_B$. 

The results depicted by the dashed and solid lines in Fig.~\ref{cm1} 
correspond to the two different procedures proposed to cure the spuriosity 
due to the lack orthogonality between initial and final states in the model. 
In the dashed line the spurious contribution is subtracted, in the solid line 
orthogonality between the s.p. states is restored. With respect to the previous 
result, shown by the dotted line, the dashed line includes all the CM effects, 
the solid line takes into account, in addition, also the effect due to the lack 
of orthogonality. It can be clearly seen from the comparison shown in the 
figure that the large difference between the old and the new results is mostly 
due to the CM effects and not to the treatment of the spuriosity or to the 
restoration  of orthogonality between the initial and final state wave 
functions of the model.  

The final cross sections given by the sum of the OB  and the two-body 
$\Delta$-currents are compared in the left panel of Fig.~\ref{cm1}. 
Calculations have been performed with the so-called $\Delta$(NN) 
parametrization \cite{sch2} for the $\Delta$-current, i.e. the parameters 
have been fixed considering the NN-scattering in the $\Delta$-region, 
where a reasonable  description of data is achieved with parameters
 similar to the ones of the full Bonn potential \cite{MaH87}.
 It was shown and explained in \cite{sch2} that in the super-parallel 
kinematics, and for the transition to the ground state of $^{14}$C, a
  regularized prescription for the $\Delta$, such as, 
e.g., $\Delta$(NN), produces a destructive interference with the OB current 
which makes the final cross section lower than the OB one. This reduction is 
strong in our previous calculation of \cite{sch2}, up to about one order of 
magnitude. 
The relevance of the destructive interference depends, however, on the 
relative weight of the OB and $\Delta$-current contributions \cite{sch2}. 
 The strong enhancement of the OB current contribution produced in the new 
calculations by CM effects reduces the 
destructive interference between the OB and $\Delta$-currents. Thus, only a 
slight reduction of the OB current contribution is obtained in the new 
calculations by the addtional incorporation of the $\Delta$-current. The final cross section is completely dominated 
by the OB current and at low values of the recoil momentum it is more than one 
order of magnitude larger than the one obtained in the old calculations. 
An enhancement factor of about 30 is given in the maximum region. 

It was shown in \cite{sch2} that dramatic differences are found in the 
super-parallel kinematics with different pa\-ra\-met\-ri\-zations of the 
$\Delta$-current and with different TOFs.

\begin{figure}
\centerline{
\resizebox{0.35\textwidth}{!}{
  \includegraphics{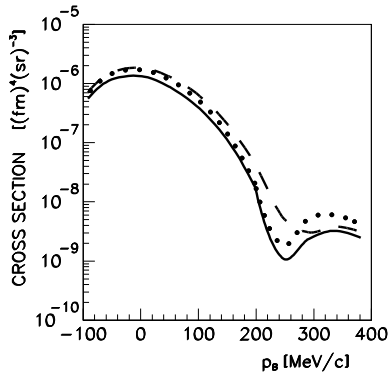}}
}
\caption{
The differential cross section of the 
$^{16}$O(e,e$'$pp)$^{14}$C$_{\mathrm{g.s.}}$  reaction as a function of  
$p_B$ in the same super-parallel kinematics as in 
Fig.~\ref{cm1}. Calculations are performed with approach C.
TOF  as in Fig.~\ref{cm1}. OB (dotted line), 
OB+$\Delta$(NoReg) (dashed line), OB+$\Delta$(NN) (solid line). 
}
\label{cm2}
\end{figure}
The cross sections calculated in approach C 
for different $\Delta$-parametrizations are shown in  
Fig.~\ref{cm2}. The result with the regularized  $\Delta$(NN) prescription,
already shown in Fig.~\ref{cm1}, is compared with the one given by the
simpler unregularized approach $\Delta$(NoReg) of \cite{sch2}. In \cite{sch2} 
the final cross sections calculated with these two parametrizations differ up 
to about one order of magnitude. Only small differences are obtained in 
Fig.~\ref{cm2}. The cross section with $\Delta$(NN) is a bit lower and the one 
with $\Delta$(NoReg) a bit higher than the cross section given by the OB 
current. 

We note that the orthogonalized wave functions are used also in the 
calculation of the matrix elements with the $\Delta$-current, where the effect 
of orthogonalization is anyhow negligible. In practice, in the present
calculations the contribution of the $\Delta$-current is the same as 
in \cite{sch2}. 
Thus, the large difference with respect to the results of \cite{sch2} in 
Figs.~\ref{cm1} and \ref{cm2} is due to 
the CM effects in the OB current and, as a consequence, to the strong 
reduction of the destructive interference between the OB and the
$\Delta$-current contribution calculated with  the $\Delta$(NN) parametrization.

\begin{figure}
\centerline{
\resizebox{0.35\textwidth}{!}{
  \includegraphics{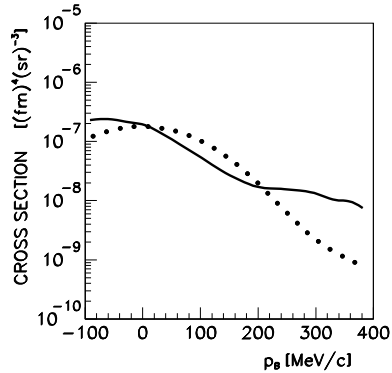}}
}
\caption{
The differential cross section of the 
$^{16}$O(e,e$'$pp)$^{14}$C$_{\mathrm{g.s.}}$  reaction as a function of 
$p_B$ in the same super-parallel kinematics as in 
Fig.~\ref{cm1}. Calculations are performed with the TOF from the simpler 
approach of \cite{GP97} and with the $\Delta$(NN) parametrization. 
The dotted line gives the result of approach A \cite{sch2}, the solid line is 
obtained with approach C.
}
\label{cm3}
\end{figure}
The cross sections shown in Fig.~\ref{cm3} are calculated with the simpler TOF
of \cite{GP97}, where the two-nucleon wave function is given 
by the product of a coupled and antisymmetrized shell model pair function and  
of a Jastrow-type central and state independent correlation function, taken 
from \cite{GD}. In this approach only SRC are considered and the final
state of the residual nucleus is a pure two-hole state. The 
ground state of $^{14}$C is a ($p_{1/2}$)$^{-2}$ hole in $^{16}$O. Thus, in 
the orthogonalized calculation the  s.p. distorted wave functions are 
orthogonalized only to the $p_{1/2}$ state.

The differences between the results of approaches A  and C, which are shown in 
Fig.~\ref{cm3}, are significant, although less dramatic than those with the TOF from the spectral function 
displayed in Fig.~\ref{cm1}, and do not change the main qualitative features of 
the previous results. It can be noted that in Fig.~\ref{cm3} the differences 
are larger at larger values of the recoil momentum, i.e. in the kinematical
region where the differences between the corresponding results in 
Fig.~\ref{cm1} are strongly reduced. 

Thus, the CM effects included in the present calculations drastically reduce 
the sensitivity to the treatment of the $\Delta$-current found in \cite{sch2} 
for the super-parallel kinematics. These CM effects are, however, very 
sensitive to the treatment of the TOF. The large differences given in the 
orthogonalized approach C by the two TOFs in Figs.~\ref{cm1} and \ref{cm3} 
confirm that the cross sections are very sensitive to the treatment of 
correlations in the TOF. This result strongly motivates further research,
 both from the experimental as well as from the theoretical side, in the
 field of pp-knockout. 

Similar calculations performed for the transition to the $1^+$ excited state of
$^{14}$C do not show any significant difference with respect to our previous
results shown in \cite{sch2}.  

The effect of the mutual interaction between the two outgoing protons (NN-FSI) 
has been neglected in the calculations presented till now because it is 
not relevant to investigate CM effects. 
NN-FSI has been studied within a perturbative treatment in \cite{sch1a,sch1}, 
where it is found that the effect depends on the kinematics and on the type of 
reaction considered. Since NN-FSI turns out to be particularly strong just  
for the $^{16}$O(e,e$'$pp)$^{14}$C$_{\mathrm{g.s.}}$ reaction and in the 
super-parallel kinematics, it can be interesting to give here only one 
numerical example for this case, just to show how our previous results 
of \cite{sch1} change in the orthogonalized approach.

\begin{figure}
\centerline{
\resizebox{0.55\textwidth}{!}{
  \includegraphics{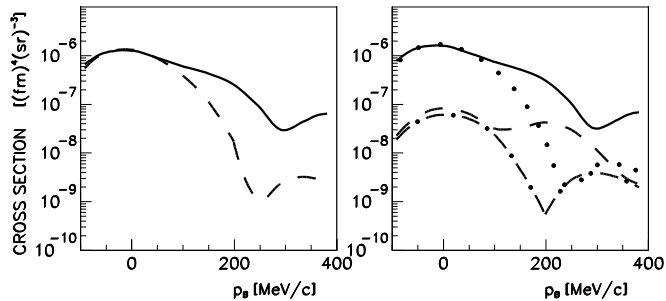}}
}
\caption{
The differential cross section of the 
$^{16}$O(e,e$'$pp)$^{14}$C$_{\mathrm{g.s.}}$  reaction as a function of  
$p_B$ in the same super-parallel kinematics as in 
Fig.~\ref{cm1}. Calculations are performed in approach C with the same TOF  and 
$\Delta$-parametrization as in Fig.~\ref{cm1}. 
Line convention for the left panel: OB+$\Delta$ with DW-NN 
(solid line), OB+$\Delta$ with DW (dashed line). Line convention for the right 
panel: OB with DW-NN (solid line), OB with DW (dotted line), $\Delta$-current 
 with DW-NN (dashed line), $\Delta$-current  with DW (dot-dashed line). 
}
\label{cm4}
\end{figure}
The effect of the NN-FSI on the cross section of the 
$^{16}$O(e,e$'$pp)$^{14}$C$_{\mathrm{g.s.}}$ reaction in the super-parallel
kinematics is shown in  Fig.~\ref{cm4}. The results obtained in the 
approach considered till now (DW), where only the interaction of each one
of the outgoing nucleons with the residual nucleus is considered, are compared
with the results of the more complete treatment (DW-NN) where also the mutual
interaction between the two outgoing nucleons is included within the same
perturbative approach as in \cite{sch1}. The cross sections given by the 
separate contributions of the OB and $\Delta$-current, as well as the ones 
given by the sum OB+$\Delta$, are displayed in the figure. These results can 
be compared with the corresponding ones presented in Figs.~3 and 4 of 
\cite{sch1}, which differ not only because the calculations of Fig.~\ref{cm4} 
are performed with orthogonalized initial and final states, but also because 
a different $\Delta$-parametrization and a different TOF are used in the two
calculations. In fact, the $\Delta$(NN) parametrization is used in 
Fig.~\ref{cm4} compared to 
  an old prescription of ours in \cite{sch1}. The TOF 
of \cite{barb} is used in  Fig.~\ref{cm4} and the one obtained from the first 
calculation of the spectral function of \cite{GPA98} in \cite{sch1}. 
The different treatment of the various theoretical ingredients produce 
significant numerical differences in the calculated cross sections. The 
contribution of NN-FSI to the final cross section is, however, of the same type 
and of about the same relevance as in \cite{sch1}. In particular, the 
considerable enhancement given by NN-FSI for medium and large values of the 
recoil momentum is confirmed in the present calculations. 
In contrast to the results of \cite{sch1}, where the enhancement at large 
$p_B$ is due the $\Delta$-current contribution, in Fig.~\ref{cm4} 
it is essentially due to the OB current, which is always dominant in the cross 
section for all the values of the recoil momentum.

\begin{figure}
\centerline{
\resizebox{0.35\textwidth}{!}{
  \includegraphics{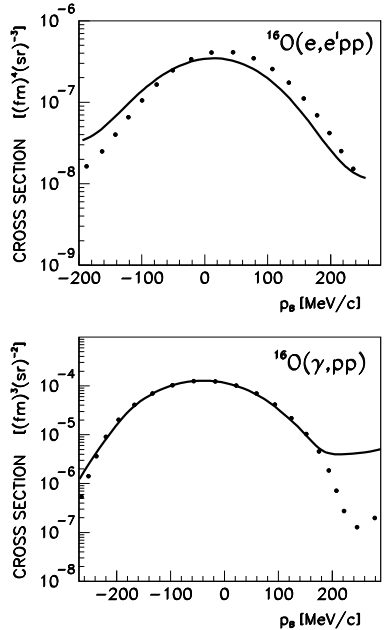}}
}
\caption{
The differential cross section of the 
$^{16}$O(e,e$'$pp)$^{14}$C$_{\mathrm{g.s.}}$ (top panel) and  
$^{16}$O($\gamma$,pp)$^{14}$C$_{\mathrm{g.s.}}$ (bottom panel) reactions as a 
function of $p_B$ in a coplanar symmetrical kinematics with $E_0= 855$ MeV, 
$\theta_e = 18^{\circ}$, $\omega=215$ MeV, and  
$q=316$ MeV/$c$ (top panel), $E_\gamma=400$ MeV (bottom panel). Different values
of $p_B$ are obtained changing the scatterimg angles of the two
outgoing protons.
Positive (negative) values of $p_B$ refer to situations where 
$\p_B$ is parallel (anti-parallel) to $\q$. Calculations are 
performed in the DW approach and with the TOF  and $\Delta$-parametrization as 
in Fig.~\ref{cm1}. 
The dotted lines are the results of approach A \cite{sch2}, the solid lines 
are obtained with approach C. 
}
\label{cm5}
\end{figure}
A different kinematical situation is considered in Fig.~ \ref{cm5}. The 
$^{16}$O(e,e$'$pp)$^{14}$C$_{\mathrm{g.s.}}$ and  
$^{16}$O($\gamma$,pp)$^{14}$C$_{\mathrm{g.s.}}$ cross sections are calculated 
in a coplanar symmetrical kinematics where the two nucleons are ejected at 
equal energies and equal but opposite angles with respect to the momentum 
transfer. 
In this kinematical setting different values of $p_B$ are obtained 
changing the scattering angles of the two outgoing protons.

The $^{16}$O(e,e$'$pp)$^{14}$C$_{\mathrm{g.s.}}$ cross section displayed in 
the top panel is calculated with $E_0= 855$ MeV, $\theta_e = 18^{\circ}$, 
and $\omega=215$ MeV, i.e. the same values as in the super-parallel kinematics. 
In this symmetrical kinematics, however, the CM effect included in the 
orthogonalized approach gives only small differences with respect to the 
previous result. The cross section is dominated by the OB current and, as 
in \cite{sch2}, it is not affected by the treatment of the $\Delta$-current. It 
is, however, sensitive to the treatment of correlations in the TOF
\cite{sch2}.

The cross section of the $^{16}$O($\gamma$,pp)$^{14}$C$_{\mathrm{g.s.}}$ 
reaction at an incident photon energy $E_\gamma=400$ MeV, displayed in the 
bottom panel of  Fig.~\ref{cm5}, is dominated by the $\Delta$-current. Thus, 
the effects due to CM and orthogonalization included in the present 
calculations, which mainly affect the OB current, do not affect the final 
cross section for recoil momentum values up to $\sim 200$ MeV/$c$. At higher 
values of  $p_B$, where the $\Delta$-current contribution is 
strongly reduced, these effects produce a large increase of the contribution 
of the OB current and of the final cross section.  
In the region where the $\Delta$-current  is dominant the sensitivity of the
results to the $\Delta$-parametrization is the same as in \cite{sch2},
i.e.\ very large. 

\section{Summary and conclusions}
\label{sec:conclusions}
Two basic aspects have been discussed within the frame of
 electromagnetic two-proton knockout reactions, i.e. CM
 effects and the spuriosity arising from the
 lacking orthogonality between initial and final state wave functions
 in connection with the usual treatments of the nuclear current. They 
 have been investigated for the cross sections of the exclusive 
 $^{16}$O(e,e$'$pp)$^{14}$C and $^{16}$O($\gamma$,pp)$^{14}$C reactions under the traditional
 conditions of super-parallel and symmetrical kinematics. 
 Different kinematics and transitions to discrete low-lying states of the residual 
 nucleus
 are known to emphasize either the role of the one-body currents, and thus
 of correlations, or of the two-body  $\Delta$-current. Since in
 two-nucleon knockout one is primarily interested in studying
 correlations, it is important to keep all the ingredients of the
 cross section under control in order to extract the useful
 information  from data.

In our previous calculations of two-nucleon knockout not all the CM effects 
were properly taken into account. In the CM frame the transition operator 
becomes a two-body operator even in the case of a one-body nuclear current. 
As a consequence, the one-body current can give a contribution to the 
cross section of two-particle emission independently of correlations. 
This  effect is similar to the one of the effective charges in
electromagnetic  reactions~\cite{ec}.

The effective transition operator entering the transition matrix
element is in principle defined consistently with the two-body initial
and final state wave  functions derived from an energy-dependent 
non-Hermitian Hamiltonian. In such an approach, no spurious
contribution comes from the orthogonality defect of the wave
functions \cite{Ox,ort}. In practice, however, one approximates the transition
operator in terms of simple forms of one- and two-body currents, 
thus introducing some spuriosity. In the past, this spuriousity was
cured by subtracting from the transition amplitude the contribution 
of the one-body current without correlations in the nuclear wave
functions. In this way, however, not only the spuriosity is
subtracted, but also the CM effect given by the two-body
operator which is present in the one-body current independently of 
correlations and which is not spurious. Alternatively, one can enforce
 orthogonality between the initial and final states by means of a
 Gram-Schmidt orthogonalization. The two approaches have been
investigated here and shown to give similar results. However, the 
Gram-Schmidt orthogonalization has been further used in the present 
investigation because it is preferable in principle and allows us to
 naturally include all the CM effects.

The CM effects due to the one-body current without correlations are different in
different situations and kinematics. For the 
$^{16}$O(e,e$'$pp)$^{14}$C$_{\rm g.s.}$ reaction in the super-parallel
kinematics these CM effects produce a strong enhancement of the contribution of
the one-body current. As a consequence, the destructive interference
 between the one-body and the two-body $\Delta$-current as well as the
 sensitivity to the treatment of the $\Delta$-current discussed in \cite{sch2} 
 are strongly reduced. With respect to the results of \cite{sch2}, the calculated
 cross section is enhanced and seems to better reproduce the experimental data
 of \cite{Rosner}. 
 On the other hand, these CM effects are very sensitive to the treatment of 
 the two-nucleon overlap function describing the initial correlated pair
 of protons. 

The mutual
interaction between the two emerging protons produces a large
enhancement of the cross section at medium and large recoil momenta. 
The effect is of the same type and of about the same relevance as in
\cite{sch1}. 
However, in contrast to the results of \cite{sch1}, where the 
enhancement at large $p_B$ is due the $\Delta$-current 
contribution, when including CM effects it is essentially 
due to the one-body current, which is always dominant in the
super-parallel cross section for all the values of the recoil
momentum.

In the symmetrical kinematics the 
$^{16}$O(e,e$'$pp)$^{14}$C$_{\rm g.s.}$ reaction is dominated by the one-body 
current and is thus sensitive to the treatment of correlations, confirming the 
result found in \cite{sch2}. In contrast, the 
$^{16}$O($\gamma$,pp)$^{14}$C$_{\rm g.s.}$ reaction is dominated by
 the $\Delta$-current and is not affected by CM and 
orthogonalization effects up to recoil momenta of the order of 200
 MeV/$c$.

In conclusion, the CM effects investigated in this work depend on kinematics
 and on the  final state of the residual nucleus. The numerical examples
 shown in the present analysis indicate that these effects are particularly
 large for the 
$^{16}$O(e,e$'$pp)$^{14}$C$_{\rm g.s.}$ reaction in the super-parallel
kinematics of the MAMI experiment. The extreme sensitivity to the treatment of
the different ingredients of the model and to different effects and
contributions makes the super-parallel kinematics very interesting but also not
particularly suitable to disentangle and investigate the specific contribution
of short-range correlations. More and different situations should be considered
to achieve this goal. Two examples have been shown in the symmetrical kinematics 
where either correlations or the $\Delta$-current are dominant. In order to
disentangle and investigate the different ingredients contributing to the   
cross sections, experimental data are needed in different kinematics which
mutually supplement each other. 

The investigation of CM effects and orthogonality between initial and final
states will be extended in a forthcoming paper to the case of electromagnetic
proton-neutron knockout,  as urgently needed after the
recent first measurements of the $^{16}$O(e,e$'$pn)$^{14}$C$_{\rm
g.s.}$ reaction performed at the MAMI microtron in Mainz \cite{middleton}.



\end{document}